\documentclass[a4paper,11pt]{article}
\usepackage{pos}

\title{Inclusive Jet Cross Section Measurements in $pp$ Collisions at $\sqrt{s} =$ 200 and 510 GeV with STAR}

\author*[a]{Zilong Chang for the STAR collaboration}

\affiliation[a]{Brookhaven National Laboratory,\\
  Upton, New York, USA 11973}

\emailAdd{zchang@bnl.gov}

\abstract{Jets, clusters of collimated particles produced in high energy proton-proton ($pp$) collisions, are an excellent tool to study the internal structure of the proton. According to perturbative QCD (pQCD) calculations, for center-of-mass energies of $\sqrt{s} = $ 200 and 510 GeV at RHIC, jet production in the mid pseudorapidity, $|\eta| <$ 1, is dominated by quark-gluon and gluon-gluon scattering processes. These jets are sensitive to gluons in the proton with momentum fraction 0.01 $<x<$ 0.5. The STAR experiment has measured a series of jet double-spin asymmetries within the pseudorapidity region of $-1 < \eta < 2$, in longitudinally polarized $pp$ collisions, to constrain the gluon helicity distribution function in the proton. Similarly, jet cross section measurements from unpolarized $pp$ collisions are effective in constraining the unpolarized gluon distribution in the proton. In this proceeding, we will present the analysis techniques and the preliminary results of inclusive jet cross section measurements in $pp$ collisions at $\sqrt{s} =$ 200 and 510 GeV.}

\FullConference{%
  *** Particles and Nuclei International Conference - PANIC2021 ***\\
  *** 5 - 10 September, 2021 ***\\
  *** Online ***
}

\begin{document}
\maketitle

\section{Introduction}
Understanding the internal structure of the proton has been an integral part of the pQCD theory. In a collinear framework, the particle production cross section can be factorized into parton distribution functions (PDFs), partonic cross sections, and fragmentation functions. The PDF is expressed as a function of the momentum fraction carried by the constituent, $x$, and the probe scale, $Q^2$.

\begin{figure}[h]
\centering
    \includegraphics[width=0.5\columnwidth]{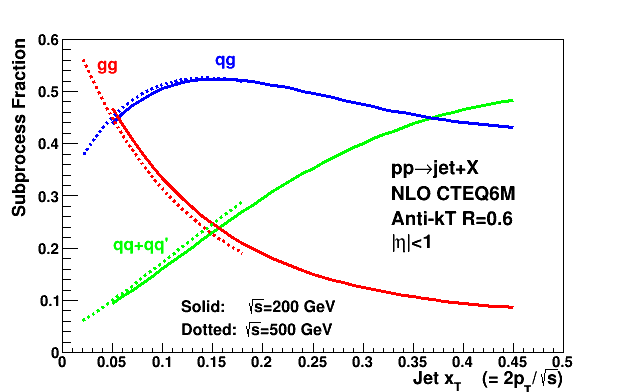}
    \caption{Fractions of jet production in $gg$ (red), $qg$ (blue) and $qq$ (green) processes as a function of jet $x_T = \frac{2p_T}{\sqrt{s}}$ at $\sqrt{s} =$ 200 and 500 GeV \cite{antikt,nlosubproc,cteq6m}.}
    \label{fig:sub}
    \end{figure}

Deep inelastic scattering (DIS) experiments have been successful in constraining the quark PDFs. Unlike DIS experiments, $pp$ collisions allow direct access to the gluon PDFs through 2 $\rightarrow$ 2 hard scatterings such as $qg$ and $gg$ processes. The final-state particles fragmented from the scattered partons can be readily detected and clustered as jets using an algorithm such as anti-$k_T$ \cite{antikt}. At center-of-mass energies $\sqrt{s} = $ 200 and 510 GeV provided by RHIC, $qg$ and $gg$ processes dominate the jet production, as shown in Fig. \ref{fig:sub} \cite{antikt,nlosubproc,cteq6m}. Therefore jet cross section measurements are sensitive to and can constrain gluon PDFs.

The STAR detector \cite{star} includes a tracking detector and two electromagnetic calorimeters covering full $2\pi$ azimuth ($\phi$) in the mid-pseudorapidity ($\eta$) region, namely the time projection chamber (TPC) in $|\eta| < 1.3$, the barrel electromagnetic calorimeter (BEMC) in $-1<\eta<1$, and the endcap electromagnetic calorimeter (EEMC) in $1.1 < \eta < 2.0$. Jet events are triggered based on the energy of a jet patch covering a region of $1\times1$ in $\eta-\phi$ in the calorimeter. The luminosity is monitored by the zero degree calorimeter (ZDC) \cite{zdc}.
\section{Luminosity}
\begin{figure}[h]
\centering
   \includegraphics[width=0.5\columnwidth]{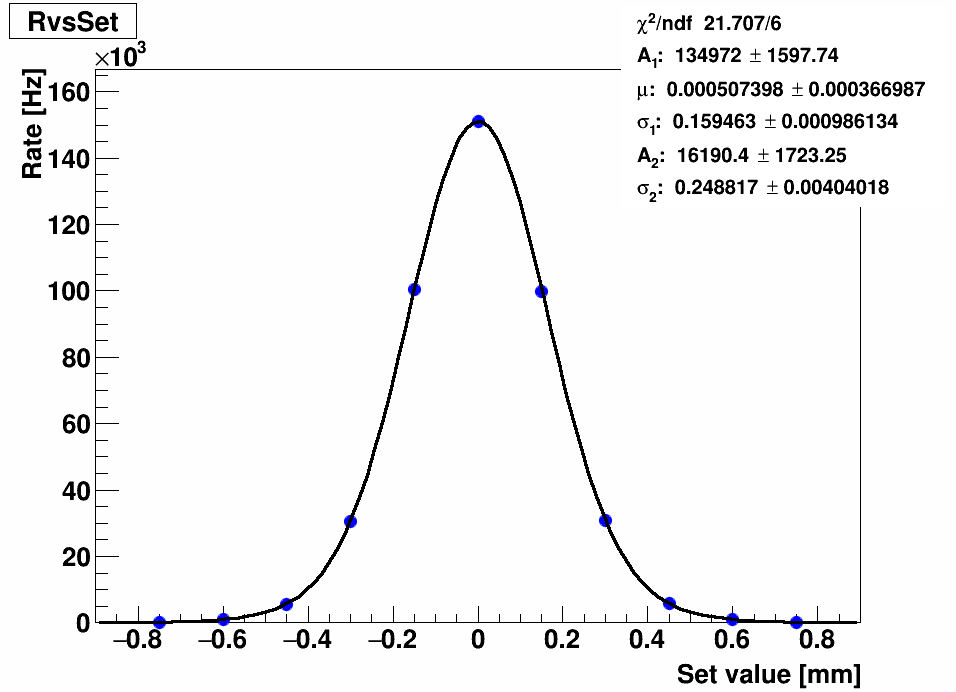}
   \caption{The corrected ZDC coincidence rate as a function of beam displacement was fitted with a double-Gaussian function.}
   \label{fig:vernier}
   \end{figure}

Given the effective beam width in the transverse plane, $h_x^{eff}$ and $h_y^{eff}$, the luminosity at the maximum overlap was expressed as $L_{0} = \frac{N_1N_2 f}{2 \pi h_x^{eff}h_y^{eff}} $, where $N_1$, and $N_2$ were number of protons in the colliding bunches of the two beams, and $f$ was the bunch crossing frequency. $h_x^{eff}$ and $h_y^{eff}$ were obtained from the change in the ZDC coincidence rate, $R_{ZDC}$, as a function of beam displacement, $x_d$, in either the horizontal or vertical direction, through the so-called vernier scans \cite{vernier}, conducted intermittently during the data taking period. The rate was corrected for accidental hits and multiple collisions \cite{amcorr} and fitted with a double Gaussian function, $R_{ZDC} = A_1e^{-\frac{1}{2}(\frac{x_d-\mu}{\sigma_1})^2}+A_2e^{-\frac{1}{2}(\frac{x_d-\mu}{\sigma_2})^2}$, where the two Gaussians shared the same mean, as shown in Fig. \ref{fig:vernier}. The effective total cross section seen by the ZDC was derived as, $\sigma_{ZDC}^{eff} = \frac{L_0}{R_{ZDC,max}}$.

By monitoring the relative ZDC coincidence rate at a specific time $t$, $R_{ZDC}(t)$, the sampled luminosity was calculated as $L = \int dt R_{ZDC}(t) \cdot \sigma_{ZDC}^{eff}$, where $R_{ZDC}(t)$ was integrated over time when the analyzed data were taken. The uncertainty on $L$ came from the statistical and systematic uncertainties associated with $\sigma_{ZDC}$ from the vernier scan. The statistical uncertainty on $\sigma_{ZDC}$ was small compared to its systematic uncertainty. The systematic uncertainty was dominated by the uncertainty of the beam displacement measurements. Other sources included beam current measurements, possible beam crossing angle, and the change in the transverse size of the beam along the longitudinal direction.
\section{Jet Reconstruction and Unfolding}
Jets were reconstructed from charged tracks measured by the TPC and energy deposits in the towers of the BEMC and EEMC using the anti-$k_T$ algorithm \cite{antikt} provided by fastjet package \cite{fastjet}. The full momentum of a charged track was subtracted from the tower energy if the track was projected onto that tower to avoid double counting. In case the energy in the projected tower was less than the track momentum, the tower energy was effectively set to zero. At $\sqrt{s} = 200$ GeV, jet parameter $R$ was chosen to be 0.6 , while $R=0.5$ was used at $\sqrt{s} = 510$. The $R$ parameter set the scale of the distance in $\eta-\phi$ space among particles in a jet. Smaller $R$ was less sensitive to more soft diffusion background created at higher $\sqrt{s}$.

An off-axis cone method was applied to the reconstructed jet transverse momentum $p_T$ to correct for the contribution from underlying events \cite{jetpaper}. This method calculated the averaged energy density, $\rho$, from the two off-axis cones centered at the same jet $\eta$ but sideways to the jet at $\pm \frac{\pi}{2}$ in $\phi$. The correction was made jet-by-jet using $dp_T = \rho \cdot A$, where $A$ was the jet area.

The reconstructed jets were unfolded to the physical particle jets by taking into account the fraction of unmatched detector jets, the response matrix, and the efficiency. The simulation was generated from the STAR tuned PYTHIA 6 event generator \cite{jetpaper}, and the GEANT3 detector simulator \cite{geant}, and embedded into zero-bias real events, which were uncorrelated with any triggers. The STAR tuned PYTHIA generator was based on the Perugia 2012 tune \cite{perugia} provided by PYTHIA 6.4.28 \cite{pythia}, with just one parameter $PARP(90)$ changed from 0.24 to 0.213. This reduced the partonic scattering amplitude at low momentum transfer, and yielded good agreement with STAR published $\pi^{\pm}$ spectra \cite{starpi1,starpi2}, especially at low $\pi^{\pm}$ $p_T$.

The number of particle jet $p_T$ bins was the same as the number of detector jet $p_T$ bins, but the bin width was slightly larger at high jet $p_T$. The unfolding process was a simple matrix inversion. The regularization was achieved by selecting the optimized number of jet $p_T$ bins and the bin width to avoid statistical fluctuations in the unfolded yields.
\section{Results}
\begin{figure}[h]
\centering
         \begin{minipage}[c]{0.4\linewidth}
    \includegraphics[width=\linewidth]{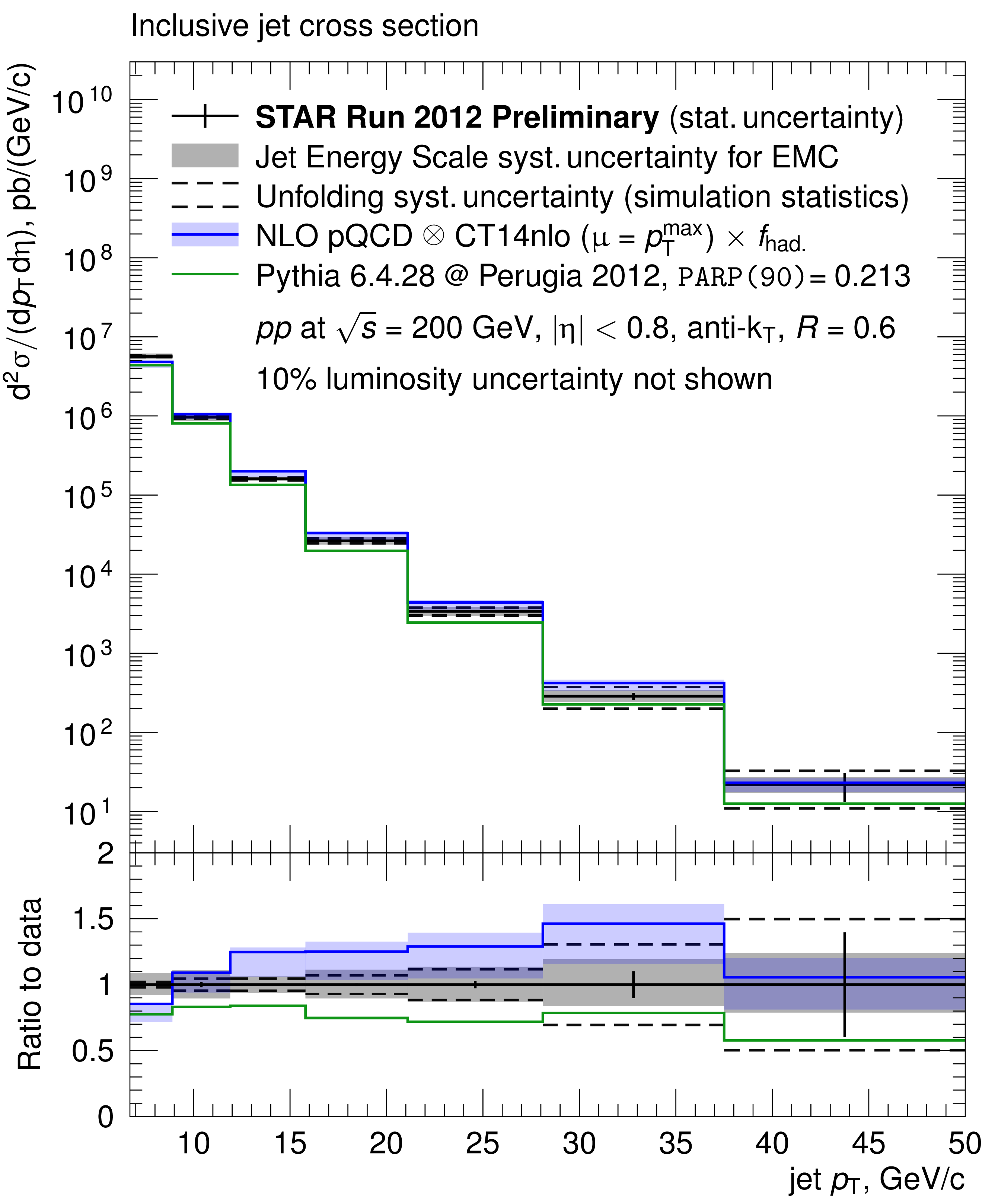}	
    \end{minipage}
    \begin{minipage}[c]{0.55\linewidth}
    \includegraphics[width=\linewidth]{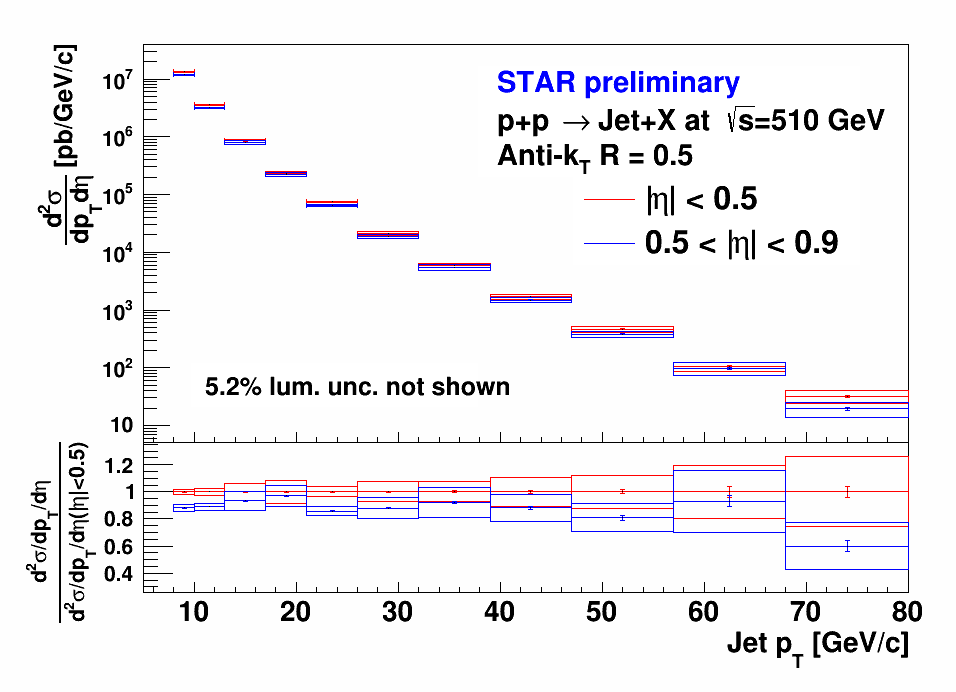}
    \end{minipage}
    \caption{Preliminary results of double differential inclusive jet cross section, $\frac{d^2\sigma}{dp_Td\eta}$, at $\sqrt{s} =$ 200 (left) and 510 (right) GeV.}
    \label{fig:crs}
    \end{figure}

The double differential inclusive jet cross section $\frac{d^2\sigma}{dp_Td\eta}$ is presented as a function of jet $p_T$ in $|\eta| <$ 0.8 at $\sqrt{s} =$ 200 GeV and in $|\eta| <$ 0.5 and $0.5<|\eta|<0.9$ at $\sqrt{s}=$ 510 GeV in Fig. \ref{fig:crs}. The systematic uncertainties include those from the electromagnetic calorimeter response to photons/electrons and hadrons, which is the dominant contribution, the TPC track momentum resolution, the TPC tracking efficiency, the unfolding bias, and the luminosity scale.
At $\sqrt{s} = 200$ GeV, the results are compared with PYTHIA 6 and pQCD calculations. The results are larger than the PYTHIA prediction across the jet $p_T$ range, but their shapes are quite close. A hadronization correction is estimated from PYTHIA. With this applied to the NLO calculation \cite{nlosubproc} at the parton level, the 200 GeV results are below the NLO predictions with the CT14 NLO PDFs \cite{ct14}.
    \section{Conclusion}
In summary, we reported the preliminary results of inclusive jet cross section at $\sqrt{s} =$ 200 and 510 GeV as a function of jet $p_T$ in different $\eta$ bins. The results will provide constraints to the unpolarized gluon PDFs at $0.01 < x < 0.5$. They will also provide normalizations for future fragmentation measurements at STAR. Complementary to data from the Tevatron and the LHC, they can be used to tune event generators, especially $\sqrt{s}$ dependent parameters. At $\sqrt{s} = 200$ GeV, the jet cross section will serve as reference data to the same measurements in Au+Au collisions at STAR in order to study the properties of the quark-gluon plasma. 

\end{document}